\documentclass[12pt,preprint]{aastex}
\shorttitle{Active Region Loops}
\shortauthors{Durgesh Tripathi et. al.}

\begin{document}

\title{On active region loops: Hinode/EIS observations}
\author{Durgesh Tripathi\thanks{Present address: }}
\affil{UCL, Mullard Space Science Laboratory, Holmbury St. Mary, Dorking, Surrey RH5 6NT, UK}
\email{d.tripathi@damtp.cam.ac.uk}
\author{Helen E. Mason}
\affil{DAMTP, University of Cambridge, Wilberforce Road, Cambridge CB3 0WA, UK}
\author{Bhola N. Dwivedi}
\affil{Department of Applied Physics, Banaras Hindu University, Varanasi-221005, India}
\author{Giulio Del Zanna\altaffilmark{1}}
\affil{UCL, Mullard Space Science Laboratory, Holmbury St. Mary, Dorking, Surrey RH5 6NT, UK}
\author{Peter R. Young}
\affil{Naval Research Laboratory, Code 7673, Washington DC 20375, USA\\
George Mason University, 4400 University Dr. MS 6A2, Fairfax, VA 22030, USA}
\altaffiltext{1}{DAMTP, University of Cambridge, Wilberforce Road, Cambridge CB3 0WA, UK}

\begin{abstract} We have carried out a study of active region loops using observations
from the Extreme-ultraviolet Imaging Spectrometer (EIS) on board
Hinode using 1~\arcsec~raster data for an active region observed on
May 19, 2007. We find that active region structures which are clearly
discernible in cooler lines ($\approx$~1MK) become 'fuzzy' at higher
temperatures ($\approx$~2MK). The active region was comprised of
red-shifted emissions (downflows) in the core and blue-shifted
emissions (upflows) at the boundary. The flow velocities estimated in
two regions located near the foot points of coronal loop showed
red-shifted emission at transition region temperature and blue shifted
emission at coronal temperature. The upflow speed in these regions
increased with temperature. For more detailed study we selected one
particular well defined loop. Downward flows are detected along the
coronal loop, being stronger in lower temperature lines (rising up to
60~km~s$^{-1}$ near the foot point). The downflow was localized
towards the footpoint in transition region lines (\ion{Mg}{7}) and
towards the loop top in high temperature line (\ion{Fe}{15}). By
carefully accounting for the background emission we found that the
loop structure was close to isothermal for each position along the
loop, with the temperature rising from around 0.8~MK to 1.5~MK from
the close to the base to higher up towards the apex
($\approx$~75Mm). We derived electron density using well established
line ratio diagnostic techniques. Electron densities along the active
region loop were found to vary from 10~$^{10}$cm$^{-3}$ close to the
footpoint to 10~$^{8.5}$cm$^{-3}$ higher up. A lower electron density,
varying from 10~$^{9}$cm$^{-3}$ close to the footpoint to
10~$^{8.5}$cm$^{-3}$ higher up, was found for the lower temperature
density diagnostic. Using these densities we derived filling factors
in along the coronal loop which can be as low as 0.02 near the base of
the loop. The filling factor increased with projected height of the
loop. These results provide important constraints on coronal loop
modeling.\end{abstract}

\section{Introduction}

Magnetically dominated solar plasma consists of a variety of
structures, such as active regions, coronal loops, X-ray bright
points, coronal holes etc. Anchored in the photosphere, coronal loops
populate both active and quiet regions, and form the basic building
blocks of the Sun's atmosphere. The physics of all kinds of loops,
therefore, holds the key to understanding coronal heating, solar wind
acceleration and the flow of mass and energy in the region. Despite
extensive work on the development of theoretical modelling \citep[see
e.g.,][for a review]{klim_review, udit} the energy source, structure
maintenance and mass balance in coronal loops are not yet fully
understood. The ultimate solution of this problem requires precise
measurements of physical plasma parameters such as flows, electron
temperature, density and filling factors in spatially resolved coronal
structures. The measurement of above mentioned parameters have been
performed using earlier spectrometers. However, the measurements were
limited by spectral and spatial resolutions. Moreover the simultaneous
temperature coverage has always been an issue.

The Extreme-ultraviolet Imaging Spectrometer \citep[EIS;][]{culhane}
on board Hinode spacecraft provides an excellent opportunity to
measure these parameters in greater detail. In this paper we have
studied overall intensity and flow structure in an active
region. Moreover, we have studied flows, electron temperature, density
and filling factor along a well resolved quiescent active region
coronal loop. The rest of the paper is structured as follows. In
section 2, we provide observations and data reduction. We present out
results in section 3 with a brief summary and conclusion in section 4.

\section{Observations and Data Preparation}

The Extreme-ultraviolet Imaging Spectrometer \citep[EIS;][]{culhane}
on board Hinode is an off-axis paraboloid design with a focal length
of 1.9~meter and mirror diameter of 15 cm. It consists of a
multi-toroidal grating which disperses the spectrum on two different
detectors covering 40~{\AA} each. The first detector covers the
wavelength range 170-210~{\AA} and second covers 250-290~{\AA}
providing observations in a broad range of temperature
($\approx$~5.8-6.7~MK). The EIS has four slit/slot options available
(1\arcsec, 2\arcsec, 40\arcsec and 266\arcsec). The EIS provides
monochromatic images of the transition region and corona at a high
cadence using a slot (wide slit). High spectral resolution images can
be obtained by rastering with a slit.

\begin{figure}
\centering
\plottwo{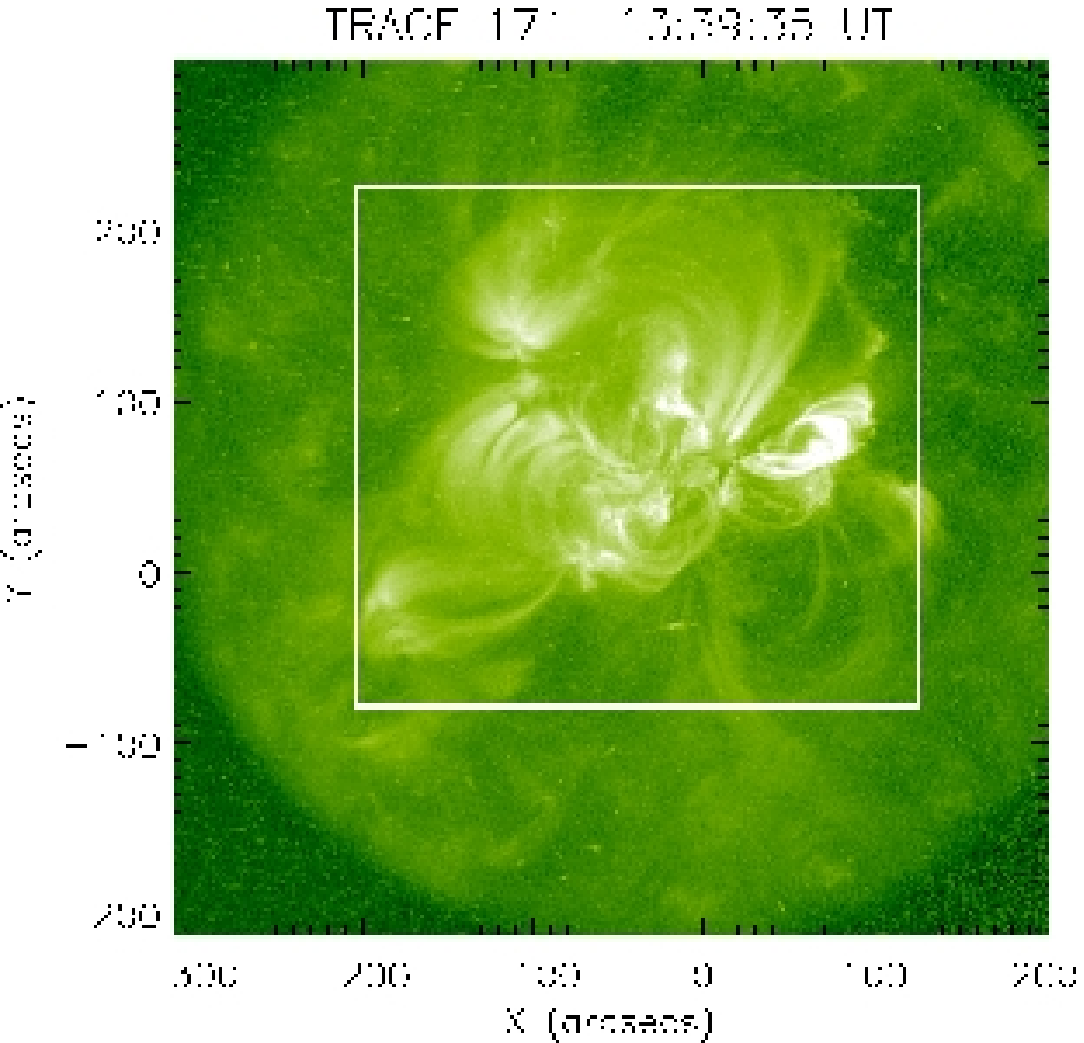}{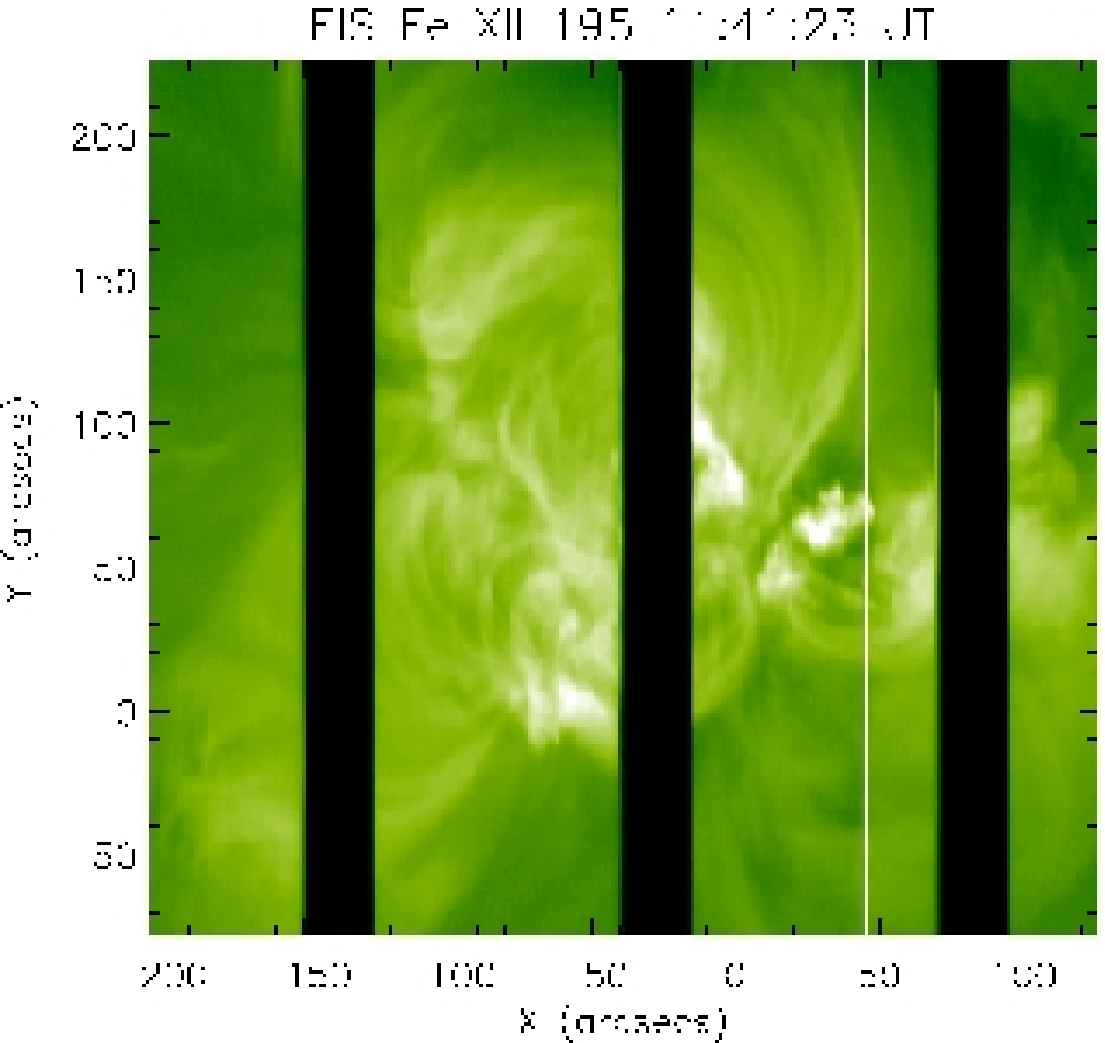}
\caption{Left panel: TRACE~$\lambda$171 image showing a snapshot of
  the active region. The white box on the image demarks the region
  which was rastered by the EIS using the 1\arcsec slit. Right panel: EIS
  image in \ion{Fe}{12}~$\lambda$195. The black stripes represent the
  data gaps due to Hinode eclipses. The white-straight line shows the
  location of the slit at 13:39:12~UT.\label{context}}
\end{figure}

The EIS observed an active region on the Sun's disk center on May 19,
2007 using the observation sequence {\it AR\_velocity\_map}. This
sequence uses the 1\arcsec slit with an exposure time of 40~seconds
with a delay of 12~seconds. The EIS raster used in this analysis
started at 11:41:23~UT and finished at 16:35:01~UT. The left panel in
Fig.~\ref{context} displays an image recorded by the Transition Region
and Coronal Explorer \citep[TRACE;][]{trace} in its 171~{\AA}
pass-band showing the complete active region taken at 13:39:35~UT. The
over-plotted box on the TRACE image represents the field-of-view (FOV)
of the EIS raster. The right panel of Fig.~\ref{context} shows an
image recorded by the EIS in \ion{Fe}{12}~$\lambda$195. During this
observation Hinode entered into eclipse mode three times, thereby
causing data gaps. After the first eclipse, EIS observed a few well
resolved loop structures between X=[0:50] arcsec. The white line
demarks the slit location at 13:39:12~UT corresponding to the time of
TRACE image shown in the left panel.

The sequence used for this study comprises very many useful spectral
lines. For this particular study, we have only selected a few windows
with lines such as \ion{Si}{7}~$\lambda$275 (log~T[K]~=~5.8),
\ion{Mg}{7}~$\lambda$278~\&~280 (log~T[K]~=~5.8),
\ion{Fe}{8}~$\lambda$186 (log~T[K]~=~5.8), \ion{Fe}{10}~$\lambda$184
(log~T[K]~=~6.0), \ion{Si}{10}~$\lambda$258~\&~261 (log~T[K]~=~6.1),
\ion{Fe}{12}~$\lambda$195~\&~186 (log~T[K]~=~6.1),
\ion{Fe}{13}~$\lambda$196~\&~203 (log~T[K]~=~6.2),
\ion{Fe}{14}~$\lambda$264~\&~274 (log~T[K]~=~6.3) and
\ion{Fe}{15}~$\lambda$284 (log~T[K]~=~6.4). The selection of these
lines provides diagnostics for the active region in a broad range of
temperatures. The data were first processed using the standard
processing routine \textit{eis\_prep.pro} provided in \textit{Solar
Software (SSW)} tree. We have fitted all the lines at each pixel of
the EIS raster using the routine \textit{eis\_auto\_fit.pro} which is
also provided in \textit{SSW}. Note that the \ion{Mg}{7}~$\lambda$278,
and \ion{Fe}{14}~$\lambda$274 are blended with a \ion{Si}{7}
line. This blending issue is taken care of by using the branching
ratios with \ion{Si}{7}~$\lambda$275. The \ion{Fe}{13}~$\lambda$203 is
blended with a \ion{Fe}{12} line which is taken care of by fitting two
Gaussians. The two \ion{Fe}{12} lines $\lambda$195 and $\lambda$186
are self blended and care must be taken while determining electron
density using these line ratios \citep[see][]{delzanna_fe12, peter2}.

It has been pointed out that images obtained from the two EIS
detectors have an offset of about 20$\arcsec$ in Y (N-S) direction and
about a couple of arcsec in X (E-W) direction. Note that the EIS slit
is oriented along N-S. Therefore a co-alignment is necessary when
comparing the outputs obtained from the two detectors. For this
purpose we have used images obtained at similar temperatures (log T[K]
= 6.1) from two CCDs (e.g., \ion{Si}{10}~$\lambda$261 \&
\ion{Fe}{12}~$\lambda$195) and co-aligned them using a
cross-correlation technique.

In order to derive physical parameters such as plasma flows,
temperature, electron density and filling factors, a number of
different spectroscopic techniques can be applied to EIS
observations. For a review on different spectroscopic techniques see
e.g. \cite{dere_mason, mason94}.

The intensity of an optically thin emission line can be written as 

\begin{equation}
I = A(z) \int_{T\rm{e}} G(T_{\rm{e}}, N_{\rm{e}})~DEM(T_{\rm{e}})~dT_{\rm{e}}
\end{equation}

where A(z) is the elemental abundance, $T_{e}$ is the electron
temperature, and $N_{e}$ is the electron density. G($T_{e}$, $N_{e}$)
is the \textit{contribution function} which contains all the relevant
atomic parameters for each line, in particular the ionization fraction
and excitation parameters. In quiet solar conditions it is usually
assumed that ionization equilibrium holds and the G($T_{e}$, $N_{e}$)
can be obtained using equilibrium ionization balance calculations. For
this purpose we have used CHIANTI, v5.2 \citep{dere97,
landi06}. $DEM(T_{e})$ is known as \textit{differential emission
measure} which is defined as:

\begin{equation}
DEM(T_{\rm{e}}) =  N_{\rm{e}}^{2}~(dh/dT_{\rm{e}}) 
\end{equation}

where dh is an element of column height along the line-of-sight (LOS).

To determine the average temperature along the LOS for a given point
along the loop we have used the \textit{EM-loci} method
\citep[see][and references therein]{delzanna2002}. In this method we
plot the ratios of observed intensities with the contribution function
I$_{obs}$/(A(z)~G(N$_{e}$,~T$_{e}$)) obtained using ionization balance
calculations as a function of temperature. If the plasma is isothermal
along the LOS then all the curves would cross at a single location
indicating a single temperature. More precisely, this method is useful
to find out if the plasma along the LOS is isothermal or
multi-thermal.

The EIS wavelength bands were chosen to cover many density diagnostic
lines \citep[see e.g.][]{delzanna2005, peter1, tripathi_ar,
peter3}. We provide measurements of average electron density along the
LOS using line ratios simultaneously from different line pairs. Once
the electron densities have been estimated, the plasma filling factor
($\phi$) - yet another important quantity - can be determined as
\citep{porter, cargill_97}

\begin{equation}
\phi = EM/N_{\rm{e}}^{2}~ h
\end{equation}

where EM is the total emission measure which is calculated by
integrating over the width of the contribution function for a given
spectral line, and h is the thickness of the loop along the LOS.

Obtaining the filling factor provides an indication of whether the
loop structures are composed of multiple strands or not. Measuring the
filling factors is very difficult due to the accumulated errors in
estimating electron density, total emission measure (i.e., ionization
equilibrium calculations) and the column depth. The main uncertainty,
however, comes from the estimation of the column depth of the emitting
plasma.

One of the main problems when studying the plasma parameters in
coronal loops is to identify a loop which does not show substantial
changes during the observing period.  Investigating the observation
sequence taken by the TRACE, we find that the well defined loop
structures seen after the first Hinode eclipse do not change
significantly, at least within the time period when the EIS was
rastering over them. Although there was a filament eruption at around
12:56~UT, one of the legs of which was rooted just west of the active
region loops, the loops studied in this paper do not show any
significant variations which could affect our study of the physical
parameters. To be absolutely sure of this, we have also studied a
raster by the EIS taken before the eruption (start: 09:42:12, end:
10:31:24~UT). We find that the loop structure has not changed from
that raster to the one for which we have derived the physical
parameters. We did not use the earlier raster to derive the physical
parameters because the exposure time was short and therefore, we did
not have enough counts for a good diagnostic study.

Since this study provides measurements of the physical parameters
along a \textit{quiescent coronal loop}, the results obtained can be
reliably used as a benchmark against theoretical models of active
region loops.

\section{Results}
\subsection{Overall Intensity and Flow Structure of the active region}

One of the most important features of the EIS is to provide
simultaneous observations at many different temperatures with an
excellent spectral resolution ($\approx$~3~km~s$^{-1}$). This,
therefore, helps us enhance our understanding of the overall
temperature structure and plasma flows in active regions.

\begin{figure}
\centering
\includegraphics[width=0.85\textwidth]{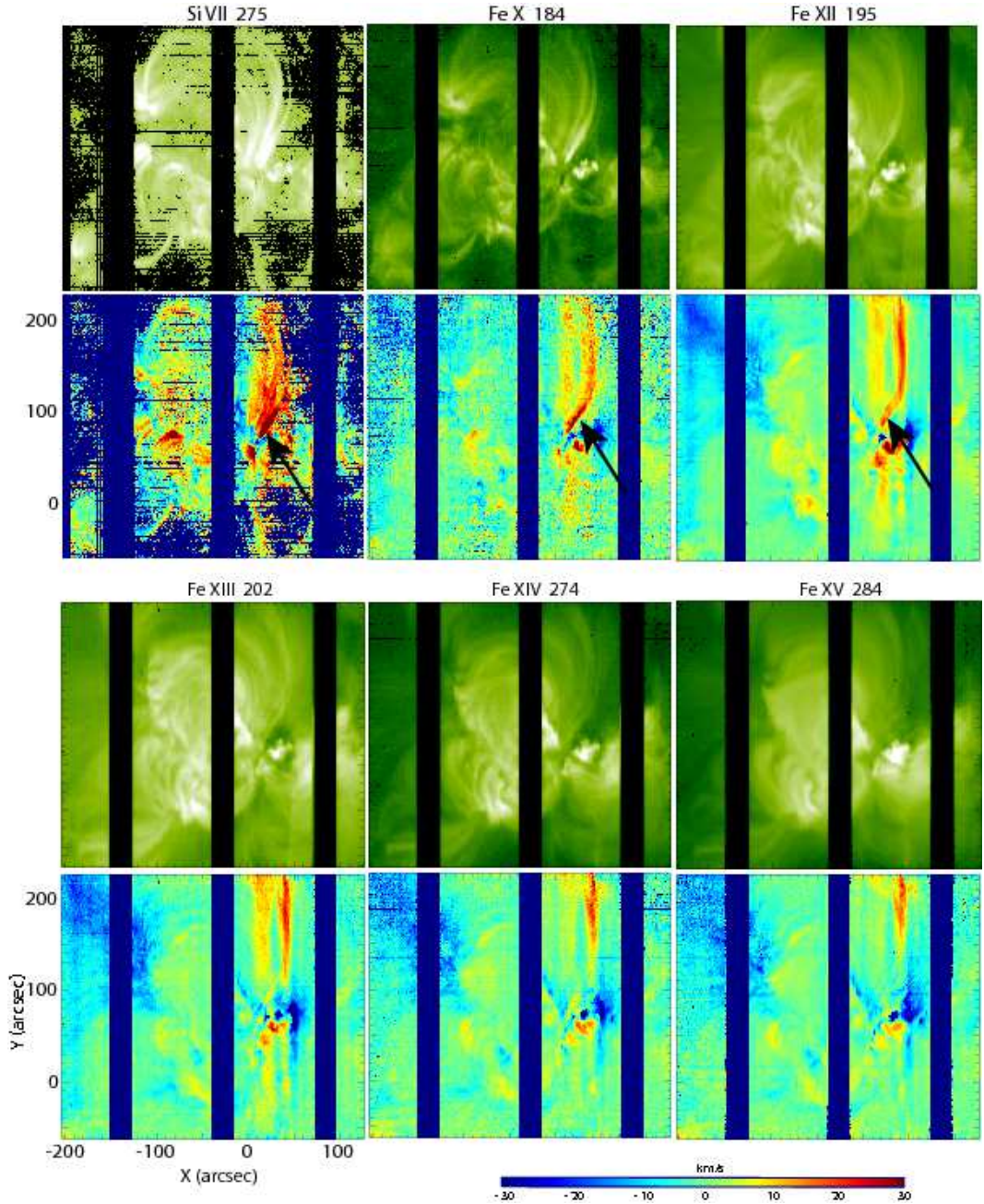}
\caption{Intensity and corresponding velocity maps obtained in
  \ion{Si}{7}, \ion{Fe}{10} and \ion{Fe}{12} (top panel) and
  \ion{Fe}{13}, \ion{Fe}{14} and \ion{Fe}{15} lines (bottom
  panel). Arrows in the top panel velocity maps locate a spatially
  resolved coronal loops.\label{int_vel}}
\end{figure}

Fig.~\ref{int_vel}, displays intensity (first and third rows) and
corresponding velocity maps (second and fourth rows) taken
simultaneously at different wavelengths. The three black stripes in
the intensity and velocity images are data gaps due to Hinode
eclipse. It is evident from the intensity images (top row) that
high-lying coronal loop structures are more prominent at cooler
temperatures. The core of the active region is seen in high
temperature lines. The active region shape become more compact and
bright at temperatures such as log~T[K]=6.3 (\ion{Fe}{14}) and
log~T[K]=6.4 (\ion{Fe}{15}). This implies that the core of the active
region is hotter confirming previous observations by e.g.,
\cite{mason99, milligan05, tripathi06} using data from the Coronal
Diagnostic Spectrometer \citep[CDS;][]{cds} aboard the Solar and
Heliospheric Spectrometer \citep[SoHO;][]{soho} and
\cite{tripathi_dublin, tripathi_ar} using EIS/Hinode observations.

The footpoint emission from the high-lying loops marked with arrows in
top row images in Fig.~\ref{int_vel} are most intense in low
temperature lines such as \ion{Si}{7} (top left image in
Fig.~\ref{int_vel}) (and \ion{Fe}{8}, not shown here) with little
emission in high temperature lines such as \ion{Fe}{13}, \ion{Fe}{14}
and \ion{Fe}{15}. This was also shown by \cite{peter1} using
Hinode/EIS observations. The high-lying loop structures are best
defined in the lower temperature lines such as \ion{Fe}{10}
(log~T[K]=6.0~MK). The loop structures merge into the diffuse
background emission with increasing temperature \citep[see
also][]{delzanna2003}. In hot temperature lines such as \ion{Fe}{15}
(and \ion{Fe}{16}, not shown here) only the core of the active region
remains visible.

In order to obtain the LOS velocity information (see the bottom rows
of in both panels of Fig.~\ref{int_vel}), instrumental effects such as
the EIS \textit{slit-tilt} and spacecraft \textit{orbital variation}
have to be compensated for. It is known that the EIS slits are not
vertical on the detectors \citep[e.g.,][]{peter2}. The 1\arcsec slit
has a tilt of about 1.2$\times$10$^{-6}$~{\AA}/pixel. Therefore, while
deriving velocities, it is essential to compensate for this
effect. This is done using the IDL routine
\textit{eis\_slit\_tilt.pro} provided in the SSW tree. After
compensating for the slit-tilt, we applied a correction for the
orbital variation using \textit{eis\_orbit\_spline.pro} also provided
in \textit{SSW}. Exploring a large number of datasets it has been
concluded that the spacecraft orbital variation changes from one EIS
study to another. Therefore, there is no automatic way to compensate
for this effect and it has to be estimated for each and every study
separately. To estimate the orbital variation, it is necessary to
select a quiet region in the raster. By doing so, we ensure that we do
not remove (add) any original (artifact) values from (to) the
data. For this purpose, we selected the bottom 50 pixels in the
quiet-sun region in this raster. By studying the average variations of
the line centroids, we could determine the effects of orbital
variation. Applying the above mentioned corrections, we obtained a
corrected set of data, to enable us to estimate the Doppler shift in
the coronal structures. To summarize, the 'at rest' reference
wavelength has been obtained by considering the bottom 50 pixels.

Based on SUMER measurements, \cite{hardi} and \cite{luca} found
that the spectral lines show non-zero line shifts in quiet-sun
observations. The cooler temperature lines (T~$<$~0.5~MK) showed red
shifted emission while the hotter lines (T~$>$~0.5~MK) were blue
shifted. Ideally, while investigating the Doppler shift measurements
this effect should be considered. However, we note that SUMER does not
observe spectral lines forming at temperature higher than log T[MK] =
6.1 i.e. temperature corresponding to \ion{Fe}{12} line. Therefore,
this effect can only be taken in account for the line such as
\ion{Si}{7}, \ion{Fe}{10} and \ion{Fe}{12} in our analysis and not for
high temperature lines studied in this paper such as \ion{Fe}{13},
\ion{Fe}{14} and \ion{Fe}{15}. Since we are interested in comparing
the flows observed in these lines with respect to each other, it is
not worthwhile taking such effects into account for some lines but not
for higher temperature lines.

The bottom row images in Fig.~\ref{int_vel} show the derived Doppler
velocity maps for different lines. The velocities maps are plotted in
the range of $\pm$~30~km~s$^{-1}$. The velocity maps show redshifts
and blueshifts all over the active region. The redshifts (downflows)
are more prominent in the core region of the active region and blue
shifts (up flows) are more prominent at the boundary of the active
region similar to the observations made by e.g.,
\cite{delzanna_dublin, delzanna_2008, harra, hara, doschek}.

The plasma flow structures in high lying loops are better defined at
lower temperatures in a similar way to structures seen in the
intensity maps. To compare the flow structure in a spatially resolved
coronal loop, we have marked a chosen loop with an arrow in the
velocity maps in top panel of Fig.~\ref{int_vel}. Plasma downflow is
seen at all temperature along the loop. However, downflows are
localized towards the foot point regions in cooler lines such as
\ion{Si}{7} and \ion{Fe}{8} (not shown in figure) and was as high as
($\approx$~60~km~s$^{-1}$). Downflows are seen all along the loop in
\ion{Fe}{10} and \ion{Fe}{12} lines being stronger towards the foot
point region. At high temperatures such as log~T[K]=6.3 (\ion{Fe}{14})
and log~T[K]=6.4 (\ion{Fe}{15}) the downflows are stronger toward the
loop top regions with almost no downflow signature at foot points. We
note that these velocities are along the LOS. Therefore the actual
velocities would be much higher than these values.

\begin{figure*}
\centering
\includegraphics[width=0.7\textwidth]{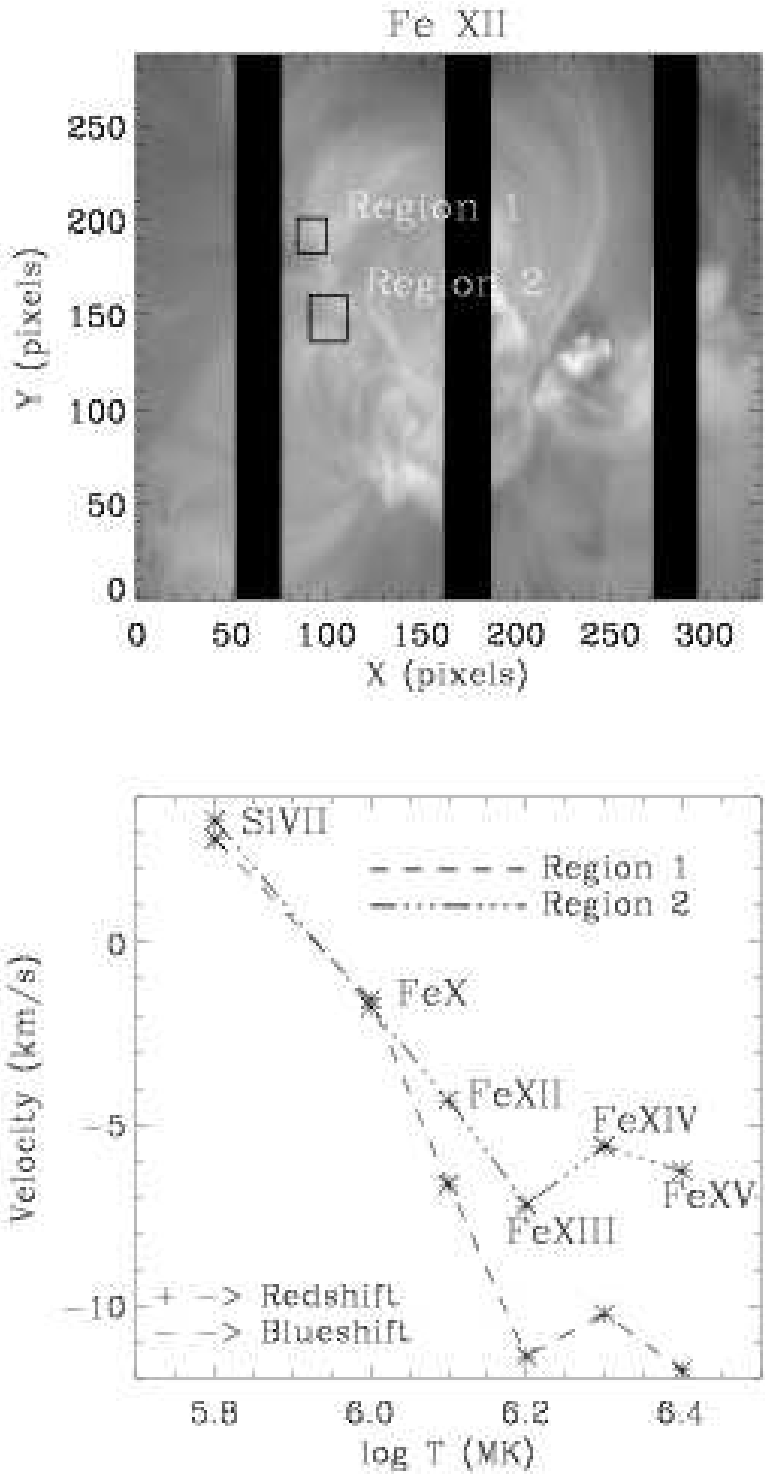}
\caption{Top panel: \ion{Fe}{12} image. Two boxes labelled Region 1
and Region 2 are chosen to measure the average velocity. Bottom panel:
Average velocities in Region 1 and Region 2 in different
lines. \label{vel_plot}}
\end{figure*}

From the intensity and velocity maps, it appears that the region
showing red-shift in cooler spectral lines shows blue-shift in high
temperature lines. To show this quantitatively we have selected two
regions labelled 'Region 1' and 'Region 2' (see top panel of
Fig.~\ref{vel_plot}) and plotted the average velocities obtained at
different lines (bottom panel of Fig.~\ref{vel_plot}). These two
regions are located at the foot points of the high lying coronal
structures. As it is evident from the plot, in \ion{Si}{7} line both
the regions show redshifted emission (downflows) turning to
blueshifted emission (upflows) with increasing temperatures. We note
that both the selected regions correspond to footpoint regions of
coronal loop. Observation of blueshifted emission in high temperature
lines at the base of coronal loops provides evidence of upflows in
loops. 

It is worth pointing out that if we take into account the results
obtained from SUMER measurements \citep{hardi, luca}, the average
downflows in \ion{Si}{7} would be weaker by about
$\approx$~3~km~s$^{-1}$ and average upflows seen in \ion{Fe}{10} and
\ion{Fe}{12} would be stronger by about $\approx$~4 and
$\approx$~9~km~s$^{-1}$ respectively. However, we do not have
measurements for other higher temperature lines from SUMER to make
such a comparison.

\subsection{Intensity variation across loop structures}

After the first Hinode eclipse, the EIS recorded a few very well
defined vertical loop structures (see Fig.~\ref{context} right panel
X=[0:50] arcsec) before going into another eclipse. The orientations
of these loops (almost North-South) are such that rastering across the
loops does not take a long time (less than ~20~mins). Therefore, this
loop system offers us an excellent opportunity to study the physical
plasma parameters along active region loops at a given time. From now
on we will only focus on the region $\approx$~X=[-10:65] arcsec and
$\approx$~Y=[70:220]. Note that the EIS rasters a region on the Sun
from West to East.

\begin{figure*}
\centering
\includegraphics[width=0.95\textwidth]{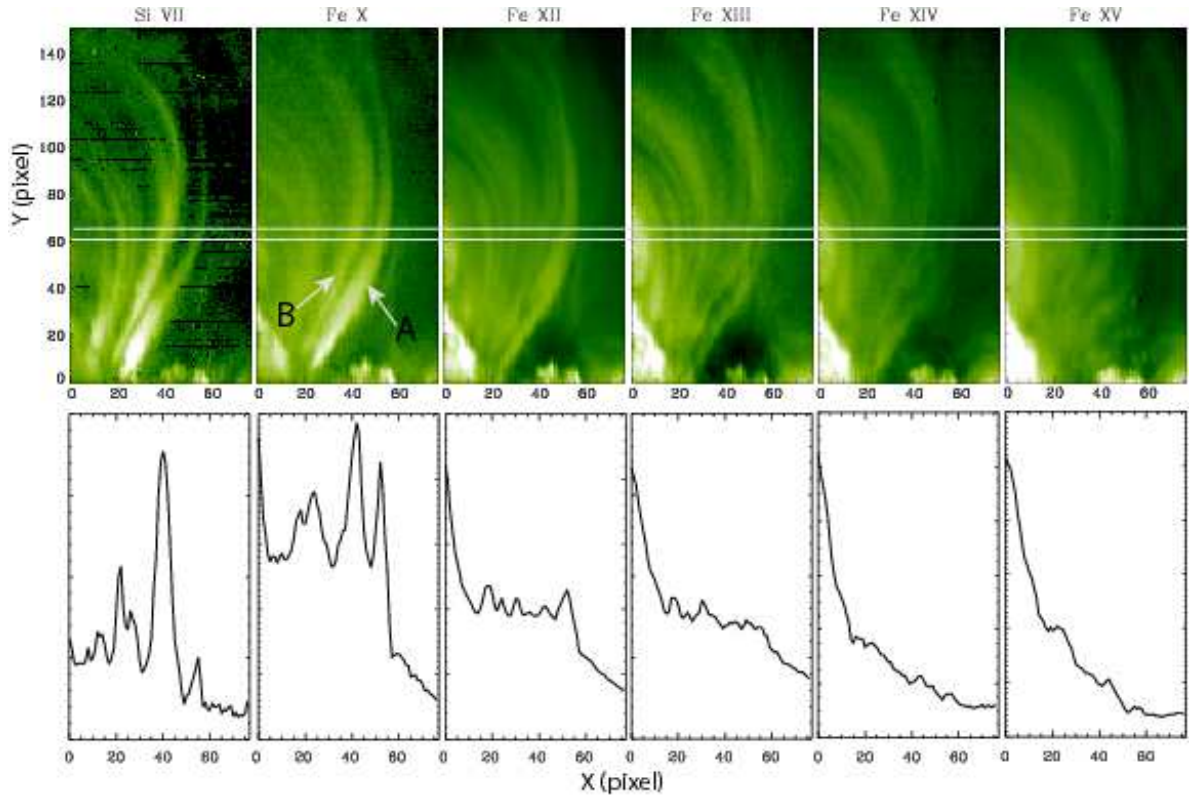}
\caption{Top panel: The active region loops seen at different
temperatures. Bottom panels: total intensity variation in the region
marked by two white lines in the corresponding top
panels.\label{loop_difftemp}}
\end{figure*}

Fig.~\ref{loop_difftemp} (top panel) shows monochromatic images of the
sub-region from the EIS raster. The spectral lines for which these
images were created are labelled for each frame. As is evident from
the figure, the coronal loops in the active region are more sharply
defined in the images observed at lower temperature lines such as
\ion{Si}{7}, \ion{Fe}{8} and \ion{Fe}{10} than those in the higher
temperature lines such as \ion{Fe}{12}, \ion{Fe}{13}, \ion{Fe}{14} and
\ion{Fe}{15}. Structures in \ion{Si}{7} are the sharpest and are
fuzziest in \ion{Fe}{15}. The structures seen in \ion{Fe}{10} are
similar to those seen in the TRACE 171~{\AA} image (see
Fig.~\ref{context} left panel). However the small difference is due to
somewhat better spatial resolution of TRACE (1\arcsec) than that of
EIS (3-4 \arcsec).

The active region loop structures are clearly discernible in the
\ion{Fe}{12} line. The images obtained in the \ion{Fe}{14} and
\ion{Fe}{15} lines lack clarity while noting that the loops seen in
\ion{Fe}{12} line may have partial overlap with \ion{Fe}{15}
line. Based on these observations it is plausible to conclude that the
coronal structures are fuzzy at higher temperatures. This in itself
has been a matter of debate amongst the solar physics community for
many years (since Skylab days). The question concerns whether the
solar corona really is fuzzy at high temperatures or simply appears to
be fuzzy because the instrumental spatial resolution and line
intensities are insufficient at higher temperatures. The high
temperature corona was observed with the Soft X-ray Telescope
\citep[SXT;][]{sxt} aboard Yohkoh, however cooler temperature emission
could not be observed with Yohkoh. The SoHO/CDS instrument could
observe the corona simultaneously at low temperature
($\approx$log~T[K]~=~5.0~MK) as well as high temperatures
$\approx$~log~T[K]~=~6.4~MK. However, the CDS spatial resolution
($\approx$~5-6~\arcsec) was not enough to unambiguously resolve this
issue. Similarly with SoHO/EIT which could observe the Sun at low
temperature using its \ion{Fe}{9} channel and at high temperature
using the \ion{Fe}{15} channel but didn't have sufficiently good
spatial resolution.

To clarify this point, we chose a region (marked between the two
horizontal white lines in the top panels of Fig.~\ref{loop_difftemp})
and plotted the corresponding intensity profile (summed in the
Y-direction) in bottom panels of Fig.~\ref{loop_difftemp}. It is
straightforward to see that the intensity fluctuations due to the
presence of spatially resolved coronal features are highly pronounced
in low temperature emission, but much less obvious at higher
temperatures. The intensity images in \ion{Si}{7} and \ion{Fe}{10}
lines, shows a spike at location X=40 which is basically due to a
spatially resolved loop, marked by an arrow as loop 'B'. Another
active region loop 'A' is also marked by arrows in the \ion{Fe}{10}
image. For both these loops we can identify enhancements in the
intensity profiles for lines such as \ion{Si}{7}, and
\ion{Fe}{10}. However in the \ion{Fe}{12} image, the intensity
corresponding to loop 'B' has almost disappeared in the fuzzy
background, but the intensity enhancement corresponding to loop 'A' is
still clearly discernible. The intensity fluctuation corresponding to
the loop 'A' is still just about discernible in \ion{Fe}{13} and
\ion{Fe}{14}. However \ion{Fe}{15} shows a sharp decrease in intensity
corresponding to the position of loop 'A'. It is interesting to note
that the intensity profile corresponding to the \ion{Fe}{15} line
shows a dip for both the coronal loops marked 'A' and 'B'. However, it
also shows a sharp enhancement (barely identifiable in lower panel
plots) between these two dips where the intensity profiles
corresponding to cooler lines show a dip. The plot also
demonstrate the possible partial overlaps between the loop structures
seen in \ion{Fe}{12} line and those seen in \ion{Fe}{15} line. This
basically implies that active regions are comprised of coronal loops
seen at different temperatures and intermingled with each other
\cite[see also][]{delzanna2006}.

\subsection{Temperature structure along Loops: Isothermal or multi-thermal?}

One of the most important issues when studying the problem of active
region heating and the coronal heating in general is whether the
coronal loops are isothermal or multi-thermal along the LOS. This
issue has been studied extensively using data from TRACE \citep[see
e.g.,][]{asch2001}, and CDS \citep[e.g.,][]{schmelz2001,
delzanna2003, schmelz2007}. The main controversy started when using the
spectroscopic techniques such as DEM, \cite{schmelz2001} showed that
the loops were not isothermal either along the loop length or along
the LOS. Later on using an observation which was taken by TRACE and
CDS simultaneously \cite{delzanna_letter} and \cite{delzanna2003}
\citep[see also][]{cirtain} showed that the loops were isothermal
along the LOS and the proper treatment of the diffuse background is
important.

Given the excellent spatial resolution of EIS and the broad
temperature coverage, we have estimated temperature along the LOS at
many given points along the loop length. For this purpose we have
applied the \textit{EM-loci} method using a set of spectral lines
which are almost density insensitive. In the \textit{EM-loci} method
we plot $I_{obs}$/(A(b)~G(N$_{e}$,T$_{e}$)) as a function of
temperature. The \textit{contribution function} G(N$_{e}$, T$_{e}$)
was calculated with CHIANTI, v.5.2, using coronal abundance of
\cite{feldman} with a density of 10$^9$~cm$^{-3}$ and the ionization
fraction of \cite{arnaud}. Similar results were obtained using
the ionization fraction of \cite{mazzotta}.

\begin{figure}
\centering
\includegraphics[width=0.6\textwidth]{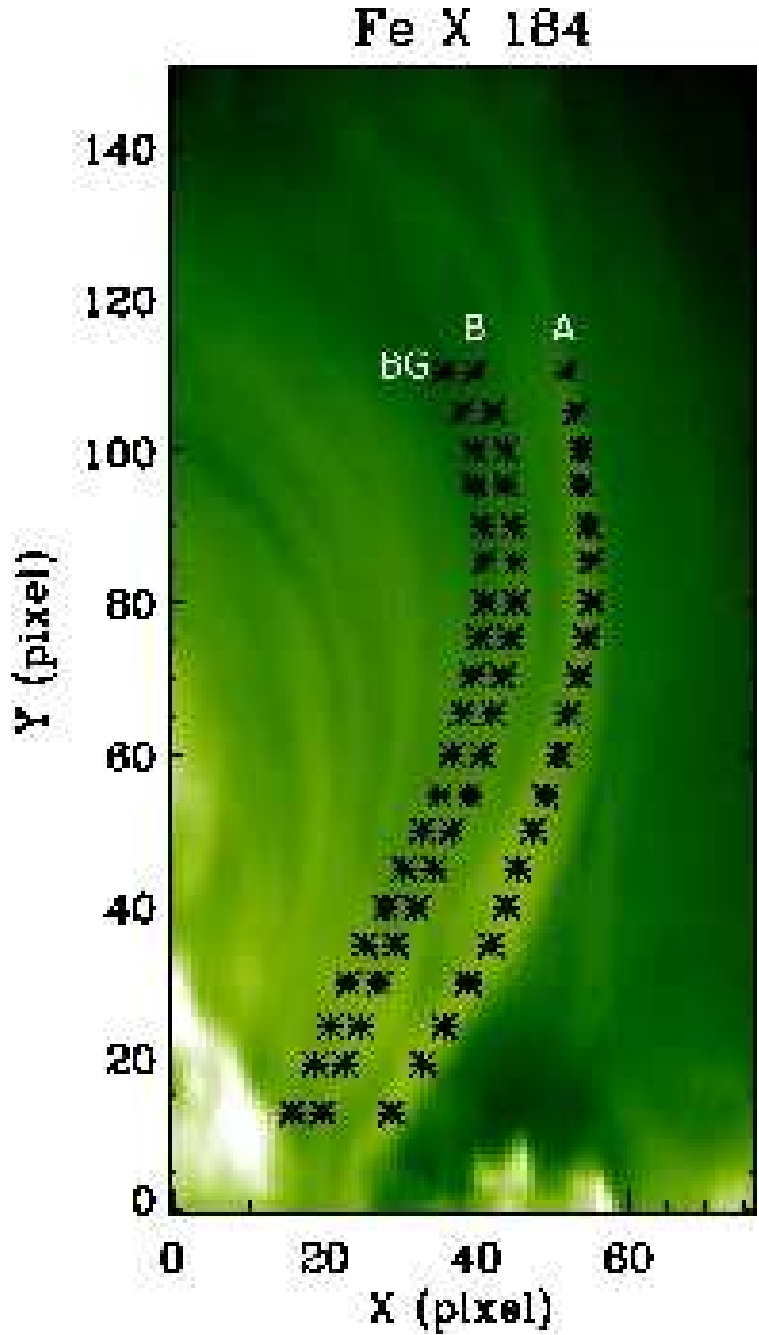}
\caption{Image obtained in \ion{Fe}{10}~$\lambda$184. Overplotted
  asterisks show two loops and the background. The right most is loop
  A and the middle is loop B. The left most one, labelled as BG, is
  used for background subtraction. \label{image_datapoints}}
\end{figure}

It is evident from Fig.~\ref{loop_difftemp} that the loops are not
sharply defined at all temperatures. They are best seen at lower
temperatures. Based on a visual inspection, therefore, we have chosen
two loops namely loop A and loop B (labelled in
Fig.\ref{image_datapoints}) in \ion{Fe}{10} and used these as a proxy
for other lines where the loops are not sharply defined. For the
background subtraction we chose a loop-shaped structure running
parallel to the loop B, shifted slightly to the East and labelled
\textit{BG} in Fig.~\ref{image_datapoints}. We chose this location to
avoid the eruptive material which might contaminate the background
emission to the West of the active region. Furthermore we have
smoothed out the intensity along the loop and background intensity
before subtraction to avoid any local fluctuations. It is worth
mentioning here that we selected different regions for background
subtraction and the results were similar with a difference of about
5-10\%.

Fig.~\ref{temperature} displays the EM-loci plots for loop A measured
at points ( 1- 20, point 1 being at the foot-point and 20th towards
the loop top) along the loops shown in Fig.~\ref{image_datapoints},
from South to North. The EM-loci plots clearly demonstrate that the
coronal loop studied here is nearly isothermal along the LOS and the
temperature increases along the loop. Close to the foot points of the
loop the temperature is $\approx$~0.8~MK and increases to
$\approx$~1.5~MK at the projected height of about 75Mm. For the first few
points (towards the loop foot point), the EM-loci for \ion{Fe}{8} is
not consistent with other lines. The discrepancy is due a problem
with the atomic data for \ion{Fe}{8} \citep{delzanna_fe8}. A similar
trend is found for loop B.

\subsection{Electron Density and Filling Factor along coronal loops}

In order to derive electron densities along the loop we have selected
several diagnostic line pairs such as \ion{Mg}{7}~$\lambda$280 \&
$\lambda$278, \ion{Si}{10}~$\lambda$258 \& $\lambda$261 and
\ion{Fe}{12}~$\lambda$186 \& $\lambda$195 \citep{peter2}. The electron
density values were obtained using the theoretical line intensity
ratios calculated using CHIANTI, v5.2 \citep{dere97,
landi06}. However, we note that the \ion{Mg}{7}~$\lambda$278 and
\ion{Fe}{14}~$\lambda$274 lines are blended with a \ion{Si}{7} line
and therefore we have used the \ion{Si}{7}~$\lambda$275 line, which is
a strong line, to remove the blend. The two \ion{Fe}{12} lines are
self blended and care must be taken while deriving densities using
\ion{Fe}{12} line ratios.

Figure~\ref{image_datapoints} shows the intensity image obtained in
\ion{Fe}{10}~$\lambda$184 line overplotted with asterisks showing
selected pixels along the loops and the background. In order to obtain
a valid electron density for a coronal loop, it is necessary to
subtract the background intensities. It has been shown that the
results can be significantly different if a proper background
subtraction is not performed \citep[see e.g.,][]{delzanna2003}. Note
that since the intensity along loop B were not significantly different
from the background intensities, it was difficult to obtain an accurate
background subtracted intensities for the weak diagnostic lines and to
derive densities. Hence for density and filling factor estimation we
have only considered loop A.

\begin{figure*}
\centering
\includegraphics[width=0.9\textwidth]{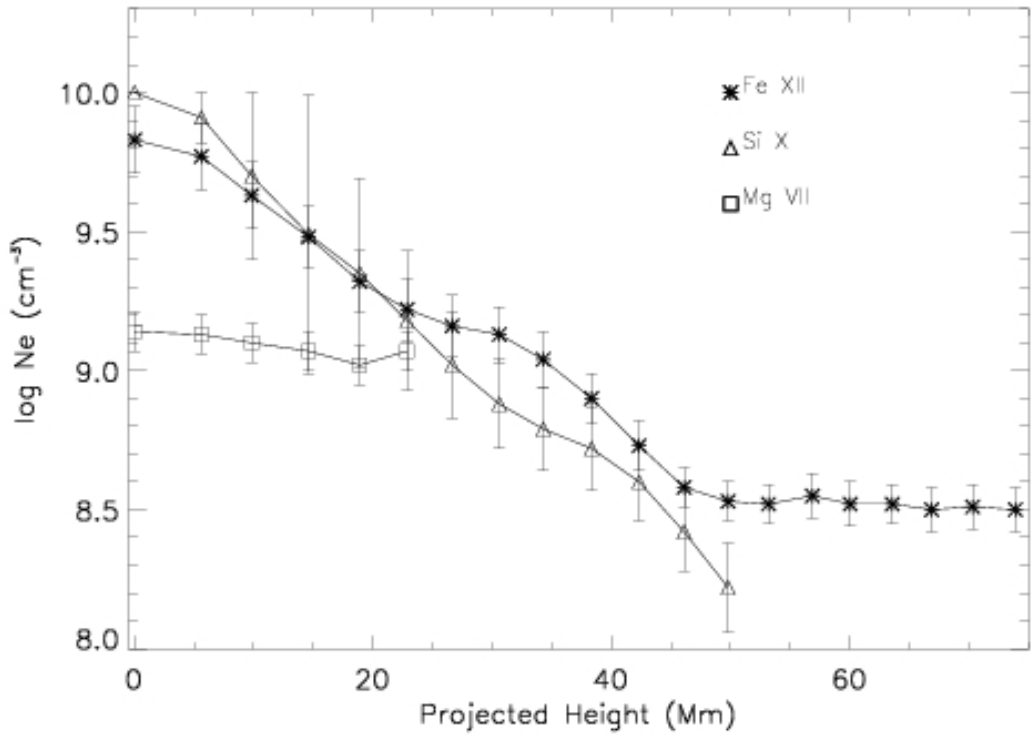}
\caption{Variation of electron density with projected height of the
loop A. Meanings of the symbols are denoted in the plot. The error bar
is estimated using 10\% error in the intensity. \label{dens_plot}}
\end{figure*}

Figure~\ref{dens_plot} shows the variation of electron density along
loop A with the projected height of the loop. The squares represent
the density values derived from \ion{Mg}{7}~$\lambda$(280/278)
(log~T[K]~=~5.8) line ratios, asterisks represent those from
\ion{Fe}{12}~$\lambda$186/195 (log~T[K]~=~6.1), and triangles are for
\ion{Si}{10}~$\lambda$258/261 (log~T[K]~=~6.1). For \ion{Mg}{7} and
\ion{Si}{10} lines we did not have enough counts at higher altitudes
along the loop and therefore, we could not estimate the density for
the higher loop locations. Error bars on densities are estimated by
considering a 10\% errors in line intensities.

It is immediately evident from Fig.~\ref{dens_plot}, that the electron
densities decrease with altitude along the loop for \ion{Fe}{12} and
\ion{Si}{10}. However for \ion{Mg}{7} it does not show as sharp a
decrease. The densities obtained from \ion{Fe}{12} and \ion{Si}{10}
are within consistent error bar. Note that for \ion{Fe}{12} the
results are based on the atomic calculation by \cite{storey}. The
electron densities obtained using \ion{Mg}{7} are significantly
different. In fact the densities using these lines could be only
measured towards the foot-point location. At the projected height of
about 20~Mm the densities using \ion{Mg}{7} become similar to those of
\ion{Fe}{12} and \ion{Si}{10}.

Another important parameter to determine is the filling factor
(defined by Eq.~3) for coronal loops. This is an indicator of whether
loop structures are composed of multiple strands or not
\citep{cargill_97}. The most problematic parameter in the filling
factor calculation is the thickness of the loop along the along the
LOS. In this study we have considered the diameter of the loop to be
equivalent to the thickness. The loop's diameter measured using
\ion{Fe}{12} image was used as thickness when estimating the filling
factors using \ion{Fe}{12} and \ion{Si}{10} lines. The loops, however,
in transition region lines such as \ion{Mg}{7} appear to be fatter
towards the footpoint. We have measured the width of the loops using
the image obtained in \ion{Mg}{7}~$\lambda$278 and used this as the
thickness along the LOS for the filling factor calculation using
\ion{Mg}{7} densities.

Table 1 provides from left to right: projected height of the loop,
thickness (diameter) of the loop along the LOS, electron densities,
filling factors and the path length for three lines such as
\ion{Fe}{12}, \ion{Si}{10}, \ion{Mg}{7}. The filling factors for
\ion{Fe}{12} densities is about 0.02 at the footpoint and about 0.8 at
a projected height of 40~Mm. For \ion{Si}{10} lines, the filling factors were
similar to those of \ion{Fe}{12} lines but become more than 1 at a
projected height of 25~Mm. We believe that this is because of the less reliable
densities obtained using \ion{Si}{10} lines ratios. We find
substantially different results for \ion{Mg}{7} with a filling factor
close to 1. For \ion{Mg}{7}, the filling factor decreases with
altitude giving reliable values only up to around 15~Mm. The filling
factor values using \ion{Mg}{7} lines were close to 1 similar to the
SoHO/CDS measurements using \ion{Mg}{7} lines
\citep[see][]{delzanna_letter, delzanna2003}.

\section{Summary and Conclusions}

The EIS on board Hinode provides an excellent opportunity to study the
physical plasma properties over a range of temperatures simultaneously
in spatially resolved coronal structures. This is extremely important
when studying the problem of active region heating.  In this paper we
have studied the overall temperature structure and flows in an active
region and in a particular spatially resolved coronal loop. In
addition we have also derived densities and filling factors along this
loop. The main results are summarized below.

\begin{itemize}

\item The high-lying loop structures are seen more distinctly in the
images obtained using lower temperature lines. With increasing
temperature the core of the active region becomes more apparent and
overall structure becomes fuzzier. The active region is comprised
throughout of upflows and downflows. The downflows are predominantly
seen in the core of the active region. However the upflows are seen at
the boundary of the active region in the low emission regions. With
increasing temperature the regions showing red shifted emission in low
temperature at the boundary of active region turns towards blue
shifted emission. (see Figs.~\ref{int_vel}, \ref{vel_plot})

\item In a well resolved coronal loop we observed downflows (red
shifts) along the loop at all temperatures. The downflows are seen
only towards the footpoint in \ion{Si}{7} and are very strong
$\approx$60~km~s$^{-1}$. However, in \ion{Fe}{10} and \ion{Fe}{12}
downflows are seen all along the loop being stronger towards the
footpoint. In high temperature lines such as \ion{Fe}{14} and
\ion{Fe}{15} the downflows are localized towards the loop top (see
Fig.~\ref{int_vel}). We note that the velocities measured at
quiet sun vary with temperature \citep{hardi}. Taking account of this
variation would not substantially affect our results.

\item The active region structures are more clearly defined in lower
temperature lines and appear 'fuzzy' at higher temperature. This is
consistent with indications from previous solar observations. However,
confirmation of this required spectroscopic observations with the same
instrument over a range of temperatures, with good spatial resolution,
as provided by the EIS (see Fig.~\ref{loop_difftemp}).

\item Using the EM-loci method we find that that both loops studied
here are almost isothermal along the LOS. The temperature at the
footpoint is about 0.80~MK and increases to about 1.5~MK at a projected height
of about 75~Mm (see Fig.~\ref{temperature}).

\item The derived electron densities decrease with loop
altitude. Towards the footpoint, the electron density estimated using
\ion{Fe}{12} ratios is log~N$_e$=9.8~cm$^{-3}$ which decreases to
log~N$_e$=8.5~cm$^{-3}$ at an altitude of about 75~Mm. The densities
obtained using \ion{Si}{10} were similar to \ion{Fe}{12}
densities. However, the estimated densities using \ion{Mg}{7} line
ratios were smaller than those obtained using \ion{Fe}{12} towards the
footpoint locations and reached a value similar to that of
\ion{Fe}{12} at a projected height of about 25~Mm (see Fig.~\ref{dens_plot}).

\item Using the densities obtained by \ion{Fe}{12}, \ion{Si}{10} and
\ion{Mg}{7} line ratios and the loop's diameter as the thickness along
the LOS, we find that the filling factor increases with projected height. The
filling factor obtained using \ion{Fe}{12} densities is about 0.02 at
the footpoint and about 0.8 at a projected height of 40~Mm. For \ion{Si}{10}
lines, the filling factors were similar to those of \ion{Fe}{12} lines
but become more than 1 at a projected height of 25~Mm. We find substantially
different results for \ion{Mg}{7} (see Table~1), closer to unity.

\end{itemize}

\section{Discussion}

An understanding of the problem the coronal heating is a long
lasting one and it is now widely believed that studying the physics of
all kinds of loops holds the key. There have been many recent
theoretical developments to explain the heating in coronal loops and
in active regions. However, the observational confirmations of these
theories have not been possible unambiguously for various reasons. For
example the 'nanoflare' model originally proposed by
\citep{parker1988} and latter on globally modelled by
\citep{cargill1994} \citep[see also][]{kopp} suggests that the coronal
loops are comprised of many sub-resolution strands and they are all
being heating separately. This predicts that the loops are
multi-thermal along the LOS. In addition, this model also predicts
high velocity evaporative upflows at high temperatures. Using the idea
of multi-stranded coronal loops in a 1D hydrodynamic simulation,
\cite{pat} suggested that these high velocity upflows can been seen as
non-thermal broadening with a particular enhancement in blue wing in
high-temperature spectral lines. Observational confirmation of these
predictions can only be unambiguously performed using spectroscopic
observations with high spectral and spatial resolution with a broad
temperature coverage.

Thanks to the excellent simultaneous spectral and spatial resolution
of EIS, we are now able to study flows and other physical parameters
in an active region and also in spatially resolved coronal structures
in detail. Observations of strong downflows at the foot point in
cooler transition region temperatures are similar to previous
observations using SUMER measurements \citep[see e.g., ][]{hardi,
luca}. In addition to that we also observed downflows only towards the
loop top in higher temperature lines such that \ion{Fe}{14} and
\ion{Fe}{15}. The observation of downflows at all temperatures raises
the question concerning the origin of the plasma in the loop and
suggests that there must be some sort of upflow of plasma filling up
the loops. \cite{pat} performed a forward modelling of \ion{Fe}{17}
emission line for EIS using data from a one dimensional hydrodynamic
simulations and suggested that the signatures of high speed upflows
can be seen as non-thermal broadening and blue-wing enhancement in the
\ion{Fe}{17} line observed by the EIS. Unfortunately, the study
sequence used in the present analysis did not contain that line with
sufficient counts preventing us to perform a direct comparison. Recently
\cite{hara} studied line profiles for \ion{Fe}{15} at foot point of
coronal loops. They observed a slight enhancement in the blue wing
suggesting that this could be associated with high velocity upflows at
higher temperature as suggested by \cite{pat}. Similar to our
observations, \cite{hara} also did not observe any signature of
downflows towards the foot point of the loops in high temperature
lines and suggested that the absence of downflows towards the
footpoint could be attributed to the presence of high velocity
upflows. Moreover, in the present study the observations of average
flow structures turning from downflows to upflows in regions located
at the foot points of the coronal loops provides another argument in
favor of upflows in coronal loops at higher temperatures. In a recent
study using the idea of nano-flare foot-point heating, \cite{patric}
predict upflows of plasma in \ion{Fe}{15} line which seem to match
with our observations.

\begin{figure}
\centering
\includegraphics[width=0.75\textwidth, angle=90]{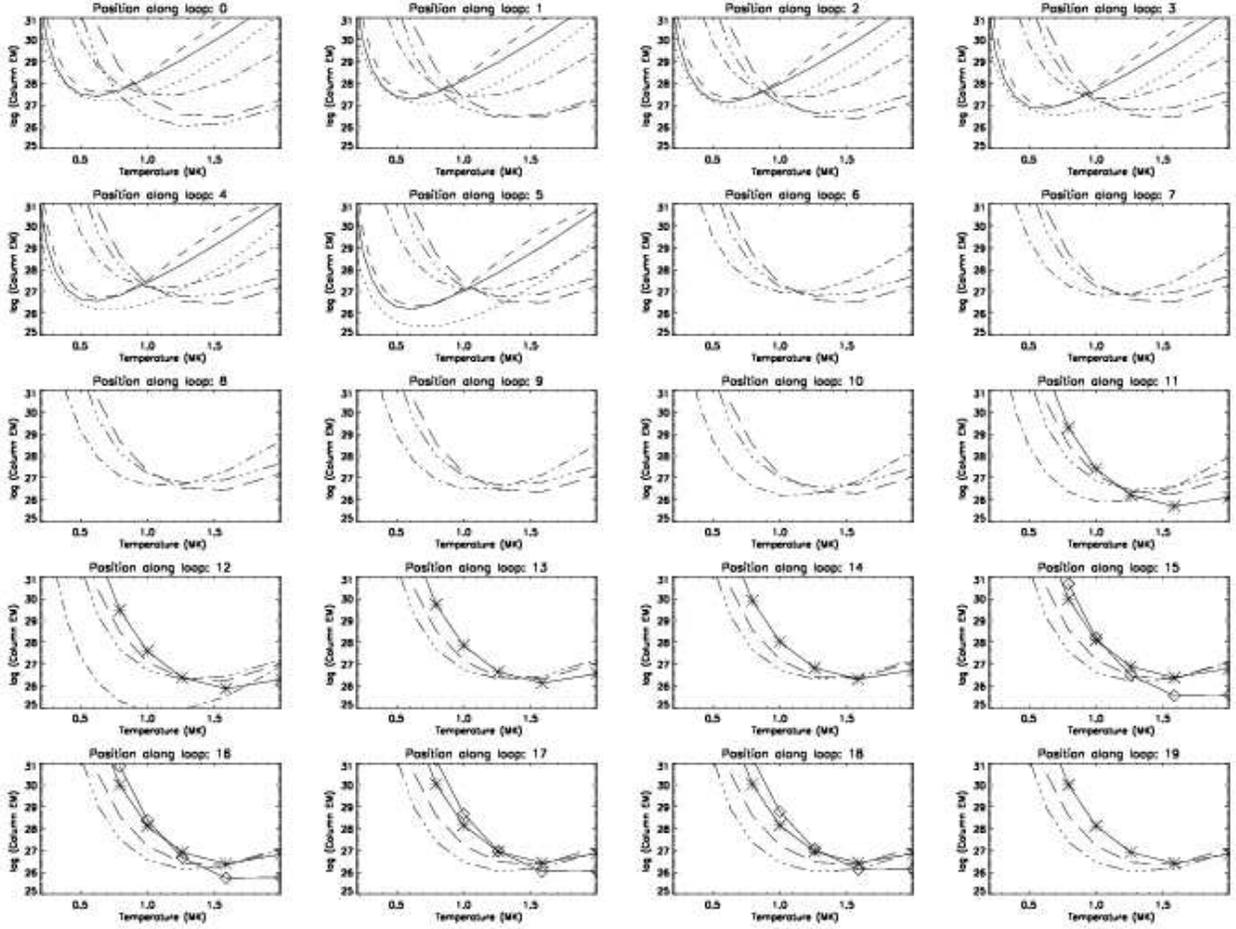}
\caption{The emission measure curves obtained from the background
subtracted line intensities for loop A. Different plots correspond to
different data points chosen along the loops. In the plots, solid
lines represent the EM-loci of \ion{Si}{7}, dotted lines \ion{Fe}{8},
dashed lines \ion{Mg}{7}, dashed-dotted \ion{Fe}{10}, dash-dot-dot-dot
\ion{Fe}{12}, and long-dashed lines represent \ion{Si}{10}. The solid
lines overplotted with astericks and diamonds represent the EM-loci of
\ion{Fe}{13} and \ion{Fe}{14} respectively.The emission measure was
calculated using \cite{arnaud} ionization fraction and coronal
abundances of \cite{feldman}.\label{temperature}}
\end{figure}

The idea of multi-stranded impulsively heated coronal loop suggests
that the filling factor values should be less than unity. This is
precisely what our measurements show. In our measurements we find that
the filling factors are much less than unity at log T[MK] = 6.1 and is
closer to unity at transition region temperature (log T[MK] = 5.8)
towards the foot point. However, with increasing projected height, the filling
factors at log T[MK] = 6.1 increases while that for log T[MK]=5.8
decreases. This means that towards the foot points of the loop most of
the plasma is at transition region temperature. However, with
increasing projected height, the amount of plasma at log T[MK] = 6.1
increases. This suggests an increasing temperature with projected height of the
loop. This is exactly what we obtained using the EM-loci analysis. The
EM-loci analysis (see Fig.~\ref{temperature}) shows that the
temperature along the loop increases with the projected height of the loop being
$\approx$0.8~MK at the foot point and $\approx$1.5~MK at a projected height of
about 75~Mm.

The EM-loci analysis also shows that the plasma along the LOS is close
to isothermal. However, the different flow structures and densities
obtained using different spectral lines and filling factors values
less than unity emphasize the multi-thermality of the plasma along the
LOS. It is not trivial to reconcile these two results together. We
note that the multi-thermality of plasma along the LOS is one of the
most important ingredient in the model of multi-stranded impulsively
heated coronal loop. However, so far we do not know how multi-thermal
or how isothermal the plasma is observationally and also
theoretically. In this analysis, e.g., at least for the first few
point from the foot point of the loop, the loops does not appear to be
exactly isothermal. There appears a narrow distribution of
temperatures.  Based on observations of a number of loops using EIS
data, \cite{warren} showed that the coronal loops which are bright at
\ion{Fe}{12} have a narrow distribution of temperature but they are
not isothermal.

The observations of high speed downflows and concurrent upflows in
coronal loops in addition to very low values of filling factors and
their nearly-isothermal structure along the LOS provide very
important constraints in the modelling of coronal loops. Given the
capabilities of EIS, more measurements should be performed for many
other coronal loops to obtain more statistically accurate values of
these parameters.

\acknowledgments

We thank the referee for constructive comments. DT, HEM and GDZ
acknowledge STFC. BND aclnowledges HEM for support through a rolling
grant from STFC at DAMTP. Hinode is a Japanese mission developed and
launched by ISAS/JAXA, collaborating with NAOJ as a domestic partner,
NASA and STFC (UK) as international partners. Scientific operation of
the Hinode mission is conducted by the Hinode science team organized
at ISAS/JAXA. This team mainly consists of scientists from institutes
in the partner countries. Support for the post-launch operation is
provided by JAXA and NAOJ (Japan), STFC (U.K.), NASA, ESA, and NSC
(Norway)

\clearpage
\begin{deluxetable}{c|rccr|ccc|cccc}
\tablecolumns{13}
\tablewidth{0pc}

\tablecaption{Derived parameters along the loop A using \ion{Fe}{12},
  \ion{Si}{10} and \ion{Mg}{7} lines. In table, 'h' is the diameter of
  the loop in arcsec, 'N$_e$' is electron density, '$\phi$' is filling
  factor and 'd' is the path length. The measured diameter was used as
  the thickness of the loop along the LOS. Note that same values of
  'h' as \ion{Fe}{12} were used while deriving filling factors using
  \ion{Si}{10} line. We have used coronal abundances of
  \cite{feldman}.}

\tablehead{
\colhead{Projected}          & \multicolumn{4}{c}{\ion{Fe}{12}}  &   \multicolumn{3}{c}{\ion{Si}{10}}   & \multicolumn{4}{c}{\ion{Mg}{7}}\\
\colhead{Height}    & \colhead{h($\arcsec$)} & \colhead{N$_e$}  &  \colhead{$\phi$}  &  \colhead{d($\arcsec$)}
		                           & \colhead{N$_e$}  &  \colhead{$\phi$}  &  \colhead{d($\arcsec$)}
                    & \colhead{h($\arcsec$)} & \colhead{N$_e$}  &  \colhead{$\phi$}  &  \colhead{d($\arcsec$)}\\
\colhead{(Mm)}      & \colhead{}           &  \colhead{(cm~s$^{-3}$)} &  \colhead{} &  \colhead{} 
		                           & \colhead{(cm~s~$^{-3}$)} &  \colhead{} &  \colhead{} 
                    & \colhead{}           &  \colhead{(cm~s$^{-3}$)} &  \colhead{} &  \colhead{}\\ 
}

\startdata
0 &  7 & 9.83 & 0.02 & 0.18 &10.00    & 0.00   & 0.02   & 20	    & 9.14    &   1.79 &  37.69\\
5 &  8 & 9.77 & 0.02 & 0.22 & 9.91    & 0.01   & 0.10   & 17 	    & 9.13    &   1.71 &  30.40\\
9 &  6 & 9.63 & 0.06 & 0.36 & 9.70    & 0.06   & 0.44   & 15 	    & 9.10    &   1.48 &  23.33\\
14 & 10 & 9.48 & 0.07 & 0.77 & 9.49    & 0.14   & 1.47   & 15 	    & 9.07    &   0.84 &  13.16\\
18 & 10 & 9.32 & 0.15 & 1.59 & 9.35    & 0.26   & 2.78   & 15	    & 9.02    &   0.56 &   8.78\\
22 & 13 & 9.22 & 0.20 & 2.69 & 9.18    & 0.43   & 5.93   & 10	    & 9.07    &   0.28 &   2.91\\
26 & 12 & 9.16 & 0.29 & 3.71 & 9.02    & 1.03   &12.99   & \nodata & \nodata &	  \nodata & \nodata\\
30 & 11 & 9.13 & 0.37 & 4.32 & 8.88    & 2.11   &24.41   & \nodata & \nodata &	  \nodata & \nodata\\
34 & 11 & 9.04 & 0.47 & 5.48 & 8.79    & 2.86   &33.08   & \nodata & \nodata &	  \nodata & \nodata\\
38 & 10 & 8.90 & 0.79 & 8.30 & 8.72    & 3.25   &34.17   & \nodata & \nodata &	  \nodata & \nodata\\
42 & 11 & 8.73 & 1.30 &15.06 & 8.60    & 4.14   &47.81   & \nodata & \nodata &	  \nodata & \nodata\\
46 & 10 & 8.58 & 2.63 &27.65 & 8.42    & 7.40   &77.62   & \nodata & \nodata &	  \nodata & \nodata\\
49 &  7 & 8.53 & 4.52 &33.19 & 8.22    & 18.73  &137.54  & \nodata & \nodata &	  \nodata & \nodata\\
53 & 12 & 8.52 & 3.11 &39.26 & \nodata & \nodata&\nodata & \nodata & \nodata &   \nodata & \nodata\\
56 & 10 & 8.55 & 3.66 &38.38 & \nodata & \nodata&\nodata & \nodata & \nodata &   \nodata & \nodata\\
60 & 10 & 8.52 & 4.21 &44.12 & \nodata & \nodata&\nodata & \nodata & \nodata &   \nodata & \nodata\\
63 &  9 & 8.52 & 4.57 &43.12 & \nodata & \nodata&\nodata & \nodata & \nodata &   \nodata & \nodata\\
66 & 11 & 8.50 & 4.41 &50.88 & \nodata & \nodata&\nodata & \nodata & \nodata &   \nodata & \nodata\\
70 & 12 & 8.51 & 3.80 &47.87 & \nodata & \nodata&\nodata & \nodata & \nodata &   \nodata & \nodata\\
73 & 12 & 8.50 & 3.89 &48.94 & \nodata & \nodata&\nodata & \nodata & \nodata &   \nodata & \nodata\\
\enddata
\end{deluxetable}

\clearpage
\clearpage
\clearpage

\clearpage
\clearpage
\clearpage
\clearpage
\clearpage

\end{document}